\documentclass[3p,times,procedia]{elsarticle}
\usepackage{nupha_ecrc}

\usepackage{color}

\usepackage[normalem]{ulem}  
\renewcommand\sout{\bgroup \color{red} \ULdepth=-.5ex \ULset}

\volume{00}

\firstpage{1}

\journalname{Nuclear Physics A}

\runauth{}

\jid{nupha}

\jnltitlelogo{Nuclear Physics A}

\usepackage{amssymb}

\usepackage[figuresright]{rotating}

\begin{document}

\begin{frontmatter}



\dochead{XXVIIIth International Conference on Ultrarelativistic Nucleus-Nucleus Collisions\\ (Quark Matter 2019)}

\title{ Number of constituent quark scaling of elliptic flows in high multiplicity p-Pb collisions at $\sqrt{s_{NN}}=$ 5.02 TeV }


\author[label1,label2]{Wenbin Zhao}
\author[label3]{Che Ming Ko}
\author[label1,label2,label6]{Yu-Xin Liu}
\author[label4,label5]{Guang-You Qin}
\author[label1,label2,label6]{Huichao Song}

\address[label1]{Department of Physics and State Key Laboratory of Nuclear Physics and Technology, Peking University, Beijing 100871, China}
\address[label2]{Collaborative Innovation Center of Quantum Matter, Beijing 100871, China}
\address[label3]{Cyclotron Institute and Department of Physics and Astronomy, Texas A$\&$M University, College Station, TX 77843, USA}
\address[label4]{Institute of Particle Physics and Key Laboratory of Quark and Lepton Physics (MOE), Central China Normal University, Wuhan, Hubei 430079, China}
\address[label5]{Nuclear Science Division, Lawrence Berkeley National Laboratory, Berkeley, CA, 94270, USA}
\address[label6]{Center for High Energy Physics, Peking University, Beijing 100871, China}

\begin{abstract}
We briefly summarize our recent study on the number of constituent quark (NCQ) scaling of hadron elliptic flows  in high multiplicity p-Pb collisions at $\sqrt{s_{NN}}=$ 5.02 TeV. With the inclusion of hadron production via the quark coalescence model at intermediate $p_T$, the viscous hydrodynamics at low $p_T$, and jet fragmentation at high $p_T$, our $Hydro-Coal-Frag$ model provides a nice description of the $p_T$-spectra and differential elliptic flow $v_2(p_T)$ of pions, kaons and protons over the $p_T$ range from 0 to 6 GeV. Our results demonstrate that including the quark coalescence is essential for reproducing the observed approximate NCQ scaling of hadron $v_2$ at intermediate $p_T$  in experiments, indicating strongly the existence of partonic degrees of freedom and the formation of quark-gluon plasma in high multiplicity p--Pb collisions at the LHC.
\end{abstract}


\end{frontmatter}
\section{Introduction}
\label{Sec1:Intro}

The strong collective flow, the number of constituent quark (NCQ) scaling of the elliptic flow, and the quenching of energetic jets are generally considered as three essential evidences for the formation of quark-gluon plasma (QGP) in relativistic heavy ion collisions~\cite{Jacobs:2007dw,long1}.  Whether these signatures can still be observed in collisions of small systems, such as p--Au, d--Au, $^3$He--Au at RHIC and p--p and p--Pb at the LHC, is a question of great current interest. Because of the limited size and lifetime of produced matter in these collisions, hard probes based on the parton energy lost in QGP can no longer leave discernible effects~\cite{Eskola:2009uj}. However,  various striking features of collectivity have recently been observed in the high-multiplicity events from collisions of  small systems~\cite{Nagle:2018nvi}. Although the color glass condensate (CGC) model based on the initial-state effect
has been shown to lead to finite azimuthal angle anisotropy~\cite{Dusling:2015gta}, the hydrodynamic model with a QGP equation of state can also successfully explain the observed collectivity in these collisions~\cite{Song:2017wtw}.
More recently, the $v_2(p_T)$ of identified hadrons have also been systematically measured in p-Pb collisions at $\sqrt{s_{NN}}$= 5.02 TeV~\cite{Pacik:2018gix}, and an approximate NCQ scaling of $v_2$ is observed at intermediate $p_T$.  This provides the possibility to further study if the partonic degrees of freedom play any cruicial roles and the QGP has been formed in collisions of small systems.  Using our newly developed hybrid model $Hydro-Coal-Frag$~\cite{Zhao:2019ehg}, we have carried out a comprehensive study on the NCQ scaling of $v_2$ in high multiplicity p-Pb collisions at $\sqrt{s_{NN}}=$ 5.02 TeV. Results from our study show that it is necessary to include the quark coalescence  contribution to hadron production in order to reproduce the observed approximate NCQ scaling of  hadron elliptic flows at intermediate $p_T$. Our study thus demonstrates the importance of partonic degrees of freedom in small collision systems and is also helpful to disentangle the origin of collectivity in high multiplicity p--Pb collisions. In the present proceedings, we briefly review our work in Ref.~\cite{Zhao:2019ehg}.

 %
\section{Methodology}
\label{Sec2:Meth}

Our study is based on the hybrid model $Hydro-Coal-Frag$ developed by us in Ref.~\cite{Zhao:2019ehg} for relativistic heavy ion collisions, which includes the production of hadrons at low $p_T$ from hydrodynamics,  at intermediate $p_T$ from quark coalescence, and at high $p_T$ from jet fragmentation. The model has been shown to simultaneously describe the $p_T$-spectra and $v_2(p_T)$ of pions, kaons and protons in high multiplicity events of p-Pb collisions over the $p_T$ range from 0 to 6 GeV~\cite{Zhao:2019ehg}. For hadron production from quark or parton coalescence, it includes thermal-thermal, thermal-hard and hard-hard parton recombinations, with the thermal partons generated from the {\tt VISH2+1} hydrodynamics~\cite{Song:2007ux} and the hard partons obtained from calculations based on the {{\tt LBT}} energy loss model~\cite{Wang:2013cia, Xing:2019xae}.  In the coalescence  calculations, mesons and baryons at intermediate $p_T$ are produced from combing quark and anti-quark pairs or three (anti-)quarks with probabilities given by the overlap of their Winger functions  with the Wigner functions of coalescing quarks~\cite{Han:2016uhh}. For the phase-space distributions of quarks and antiquarks, they are taken from the {\tt VISH2+1} hydrodynamics for soft partons with $p_T>p_{T1}$ and from the {{\tt LBT}} energy loss model for hard partons with $p_T>p_{T2}$, where $p_{T1}$ and $p_{T2}$ are tunable parameters in this model. For remnant hard partons not used for hadron production from quark coalescence, they are fragmented to hard hadrons using the ``hadron standalone mode" of PYTHIA8~\cite{Sjostrand:2007gs}.  Soft hadrons at low $p_T$ are, on the other hand, produced from the hyper-surface of {\tt VISH2+1} hydrodynamics using the Cooper-Frye formula. To avoid double counting, only  thermal mesons and baryons with $p_T$ below 2$p_{T1}$ and 3$p_{T1}$, respectively, are included. All these produced hadrons are further fed into the UrQMD model~\cite{Bass:1998ca} to take into account the subsequent evolution of the hadronic matter until kinetic freeze-out when hadrons stop scattering.

The two main tunable parameters $p_{T1}$ and $p_{T2}$ in the $Hydro-Coal-Frag$ model are fixed by fitting the $p_T$-spectra of pions, kaons and protons as well as the  $p/\pi$ yield ratio in high multiplicity p-Pb collisions at $\sqrt{s_{NN}}=$5.02 TeV.  With the resulting values $p_{T1}=1.6$ GeV, $p_{T2}=2.6$ GeV, one can then predict the differential elliptic flow $v_2$ and the related NCQ scaled $v_2$ for identified hadrons. The importance of hadron production from quark coalescence can thus be studied by comparing results from the $Hydro-Coal-Frag$ model with those from the usual {\tt Hydro-Frag} model that includes only low $p_T$ hadrons from hydrodynamics and high $p_T$ hadrons from jet fragmentation.

\section{Results and discussions}
\label{Sec3:RES}

\begin{figure*}[ht]
\centering
\includegraphics[width=0.82\textwidth]{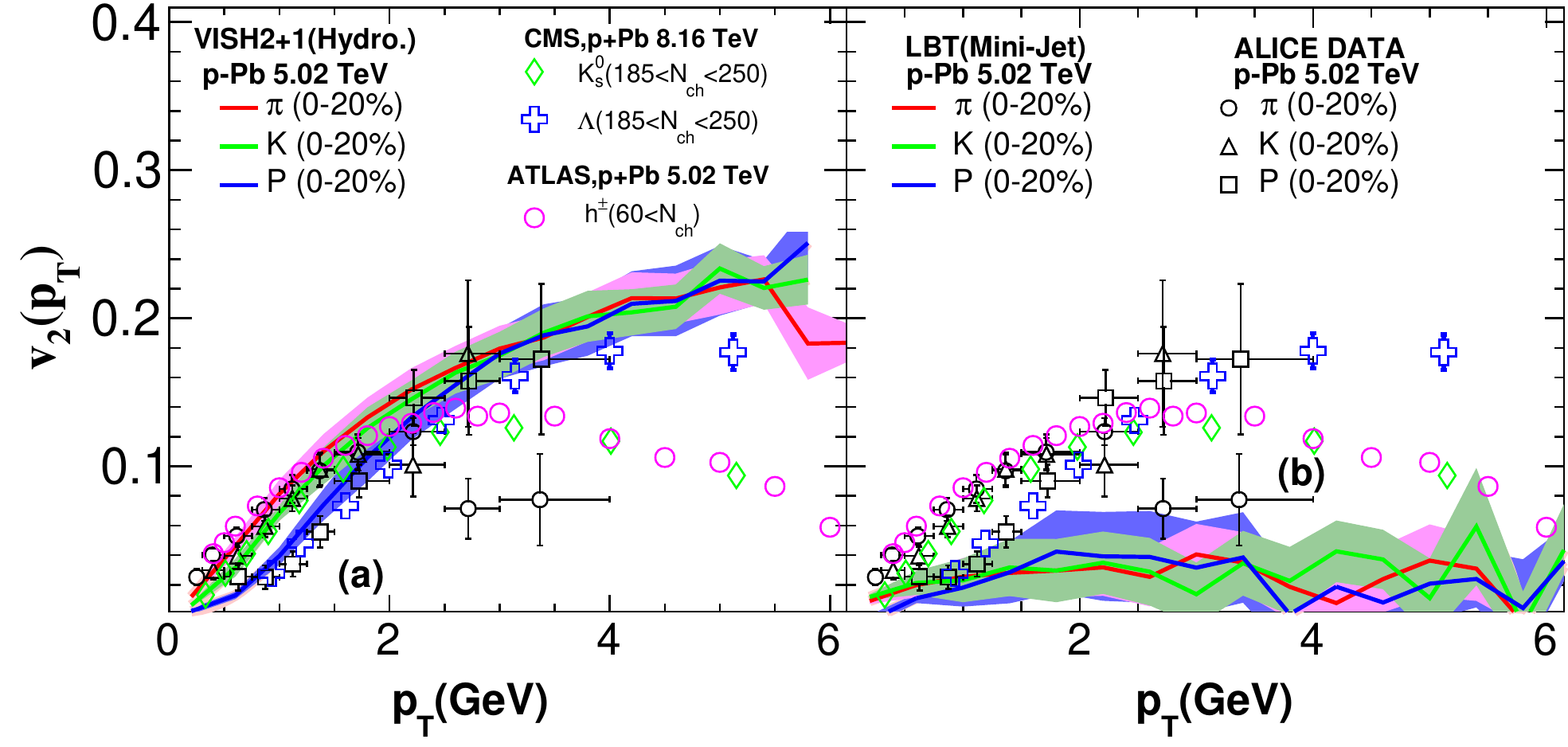}
\caption{(Color online) Differential elliptic flow $v_2(p_T)$ from {\tt VISHNU}  (a) and {\tt LBT}  (b) model calculations. Data are taken from ALICE, CMS and ATLAS collaborations~\cite{ABELEV:2013wsa, Sirunyan:2018toe,Aaboud:2016yar}. }
\label{fig:distrov2v3}
\end{figure*}

Figure~1 shows the differential elliptic flow $v_2(p_T)$ of various hadrons in 0-20\% p-Pb collisions at $\sqrt{s_{NN}}=5.02$ TeV, calculated with either the {\tt VISHNU} hydrodynamics (panel (a)) or the {\tt LBT} model with only mini-jets (panel (b)). Compared to the experimental data shown in the figure, the {\tt VISHNU} hydrodynamics describes well the data below 2 GeV but overestimates it at intermediate $p_T$\sout{,} and gives too large values at high $p_T$. For the {\tt LBT} model with mini-jets, its predicted elliptic flows are, on the other hand, significantly below the data for the whole $p_T$ range below 6 GeV.

\begin{figure*}[ht]
\centering
\includegraphics[width=0.9\textwidth]{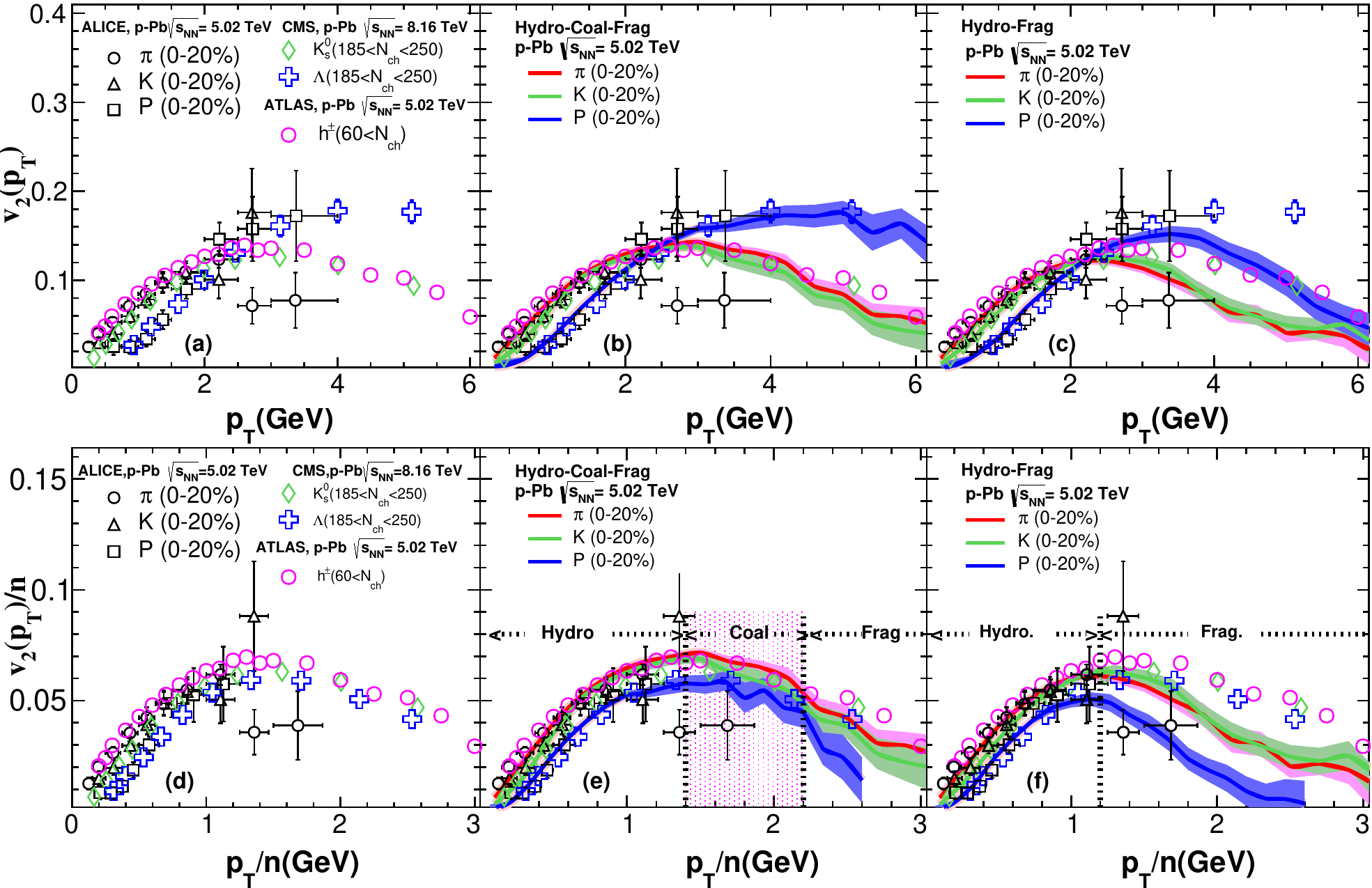}
\caption{(Color online) Differential elliptic flow $v_2(p_T)$ and the number of constituent quark scaled $v_2(p_T)$ of pions, kaons and protons in 0-20\% p-Pb  at $\sqrt{s_{NN}}=5.02$ TeV, measured in experiments (panels (a) and (d)) or calculated with the {\tt Hydro-Coal-Frag} model  (panels (c) and (e)) and the {\tt Hydro-Frag} model (panel (c) and (f)).  The ALICE, CMS and ATLAS data are taken from Refs.~\cite{ABELEV:2013wsa}, ~\cite{Sirunyan:2018toe} and~\cite{Aaboud:2016yar}, respectively. }
\label{fig:distrov2v3}
\end{figure*}

Figure~2 presents the elliptic flow $v_2(p_T)$ of pions, kaons and protons (upper panels) and their NCQ scaled elliptic flow $v_2(p_T)/n$ (lower panels) calculated  with $n=2$ for mesons and $n=3$ for baryons.  For clarity, experimental data  from ALICE~\cite{ABELEV:2013wsa}, CMS~\cite{Sirunyan:2018toe} and ATLAS~\cite{Aaboud:2016yar} Collaborations are shown in left panels, and results from the {\tt Hydro-Coal-Frag} and {\tt Hydro-Frag}
models are given in the middle and right panels, respectively. The {\tt Hydro-Coal-Frag} model, which includes hadron production from hydrodynamics, quark coalescence and jet fragmentation, is seen to describe nicely the measured elliptic flow $v_2(p_T)$ of these identified hadrons from 0 to 6 GeV.  Also, the quark coalescence contribution is important for reproducing the observed {interception} of  pion and proton $v_2(p_T)$ at $p_T$ around 2.5 GeV. In contrast, the {\tt Hydro-Frag} model without the quark coalescence contribution fails to describe the $v_2(p_T)$ of identified hadrons above 3 GeV and also gives a smaller $p_T$ at which the pion and proton $v_2(p_T)$ {intercept}, no matter how the parameters in this model are tuned.

For the NCQ scaling of hadron $v_2$, panels (d) and (e) show that the {\tt Hydro-Coal-Frag} model reproduces the approximate scaling behavior between 1.4$<p_T/n<$2.2 GeV observed in experiments. The slight violation of the NCQ scaling in the {\tt Hydro-Coal-Frag} model is due to resonance decays and jet fragmentation. On the other hand, the {\tt Hydro-Frag} model without the quark coalescence contribution to hadron production not only underestimates the magnitude of $v_2(p_T)/n$ measured in experiments by about 50\% but also leads to a violation of the NCQ scaling at intermediate $p_T$.  The success of the {\tt Hydro-Coal-Frag} model in reproducing the experimental data on the  $v_2$($p_T$) of identified hadrons and their NCQ scaling at intermediate $p_T$ in high multiplicity p-Pb collisions at the LHC thus gives a strong indication for the importance of hadron production from quark coalescence.

\section{Summary}
\label{Sec4:Sum}

We have presented here some results from our detailed study on the NCQ scaling of elliptic flow $v_2$ at intermediate $p_T$ in high multiplicity p-Pb collisions at $\sqrt{s_{NN}}=$ 5.02 TeV, based on the coalescence model that uses soft thermal partons from the {\tt VISH2+1} hydrodynamics and hard partons from the energy loss model {\tt LBT}. With low $p_T$ hadrons from the hydrodynamics, high $p_T$ hadrons from jet fragmentation and intermediate $p_T$ hadrons from quark coalescence, our {\tt Hydro-Coal-Frag} hybrid model can simultaneously describe the $p_T$-spectra and $p_T$-differential $v_2(p_T)$ of identified hadrons over the $p_T$ range from 0 to 6 GeV.  We have demonstrated that the inclusion of the quark coalescence contribution to hadron production is essential for reproducing the experimentally observed approximate NCQ scaling of $v_2$ at intermediate $p_T$.  Our results thus strongly indicate the existence of the partonic degrees of freedom and the possible formation of the QGP in high multiplicity p-Pb collisions at the LHC. \\




\noindent \textsl{Acknowledgments.} W. Z., Y. L. and H. S. are supported by the NSFC under grant Nos. 11675004 and G.-Y. Q. is supported by Natural Science Foundation of China (NSFC) under Grants No. 11775095, 11890711 and 11375072, and by the China Scholarship Council (CSC) under Grant No. 201906775042.  C.M. K. is supported by {the} US {Department of Energy} under Contract No. DE-SC0015266 and the Welch Foundation under Grant No. A-1358. W. Z. and H. S. also gratefully acknowledge the extensive computing resources provided by the Supercomputing Center of Chinese Academy of Science (SCCAS), Tianhe-1A from the National Super computing Center in Tianjin, China{,} and the High-performance Computing Platform of Peking University.

\bibliographystyle{elsarticle-num}

\begin{thebibliography}{99}


\bibitem{Jacobs:2007dw}
  P.~Jacobs, D.~Kharzeev, B.~Muller, J.~Nagle, K.~Rajagopal and S.~Vigdor,
  \emph{``Phases of QCD: Summary of the Rutgers long range plan town meeting, January 12-14, 2007,''}
  arXiv:0705.1930.

\bibitem{long1}
  \emph{``The Frontiers of Nuclear Science, A Long Range Plan,''}
  arXiv:0809.3137.


\bibitem{Nagle:2018nvi}
  J.~L.~Nagle and W.~A.~Zajc,
  Ann.\ Rev.\ Nucl.\ Part.\ Sci.\  {\bf 68}, 211 (2018).


\bibitem{Dusling:2015gta}
  K.~Dusling, W.~Li and B.~Schenke,
  Int.\ J.\ Mod.\ Phys.\ E {\bf 25}, no. 01, 1630002 (2016).

\bibitem{Song:2017wtw}
  H.~Song, Y.~Zhou and K.~Gajdosova,
  Nucl.\ Sci.\ Tech.\  {\bf 28}, no. 7, 99 (2017).




\bibitem{Eskola:2009uj}
  K.~J.~Eskola, H.~Paukkunen and C.~A.~Salgado,
  JHEP {\bf 0904}, 065 (2009).



\bibitem{Pacik:2018gix}
  V.~Pac\`ik [ALICE Collaboration],
  Nucl.\ Phys.\ A {\bf 982}, 451 (2019).



\bibitem{Zhao:2019ehg}
  W.~Zhao, C.~M.~Ko, Y.~X.~Liu, G.~Y.~Qin and H.~Song,
  arXiv:1911.00826 [nucl-th].


\bibitem{Song:2007ux}
  H.~Song and U.~W.~Heinz,
  Phys.\ Rev.\ C {\bf 77}, 064901 (2008);
  H.~Song and U.~W.~Heinz,
  Phys.\ Lett.\ B {\bf 658}, 279 (2008);
  H.~Song, S.~A.~Bass, U.~Heinz, T.~Hirano and C.~Shen,
  Phys.\ Rev.\ Lett.\  {\bf 106}, 192301 (2011).


\bibitem{Wang:2013cia}
  X.~N.~Wang and Y.~Zhu,
  Phys.\ Rev.\ Lett.\  {\bf 111}, no. 6, 062301 (2013).

\bibitem{Xing:2019xae}
  S.~Cao, T.~Luo, G.~Y.~Qin and X.~N.~Wang,
  Phys.\ Rev.\ C {\bf 94}, no. 1, 014909 (2016);


\bibitem{Han:2016uhh}
  K.~C.~Han, R.~J.~Fries and C.~M.~Ko,
  Phys.\ Rev.\ C {\bf 93}, no. 4, 045207 (2016).








\bibitem{Sjostrand:2007gs}
  T.~Sjostrand, S.~Mrenna and P.~Z.~Skands,
  Comput.\ Phys.\ Commun.\  {\bf 178}, 852 (2008).

\bibitem{Bass:1998ca}
  S.~A.~Bass {\it et al.},
  Prog.\ Part.\ Nucl.\ Phys.\  {\bf 41}, 255 (1998).


\bibitem{ABELEV:2013wsa}
  B.~B.~Abelev {\it et al.} [ALICE Collaboration],
  Phys.\ Lett.\ B {\bf 726}, 164 (2013).

\bibitem{Sirunyan:2018toe}
  A.~M.~Sirunyan {\it et al.} [CMS Collaboration],
  Phys.\ Rev.\ Lett.\  {\bf 121}, no. 8, 082301 (2018).


\bibitem{Aaboud:2016yar}
  M.~Aaboud {\it et al.} [ATLAS Collaboration],
  Phys.\ Rev.\ C {\bf 96}, no. 2, 024908 (2017).




\end{thebibliography}

\end{document}